\documentstyle[11pt]{article}
\makeatletter
\let\chapter\hid@chapter
\makeatother
\def\lsim{\lower.5ex\hbox{$\; \buildrel < \over \sim \;$}}
\def\gsim{\lower.5ex\hbox{$\; \buildrel > \over \sim \;$}}
\begin{document}
\pagenumbering{arabic}
\title{Study of Accretion Processes on Black Holes: Fifty Years of Developments}

\author{Sandip K. Chakrabarti \\
S.N. Bose National Centre for Basic Sciences\\
JD-Block, Sector-III, Salt Lake, Kolkata 700098, INDIA
\footnote{ Also at Centre for Space Physics P-61 Southend Gardens, Kolkata, 700084} \footnote{e-mail: chakraba@boson.bose.res.in}
}
\maketitle

\begin{abstract}

Fifty years ago, in 1952, the first significant paper on accretion
flows was written by Bondi. The subject has grown exponentially
since then. In fact, today many of the satellites engaged in space physics
research look for signatures of accretion processes in whatever objects 
are studied. In this review, I will touch upon the significant developments
in these years in this subject, emphasizing mainly on accretion onto
black holes. Since winds and accretions are generally studied under similar framework,
some references of the winds/outflows studies will also be made.
\end{abstract}

\noindent PUBLISHED IN `FRONTIERS IN ASTROPHYSICS' Ed. S.K. Chakrabarti, Allied Publishers, p. 145, 2002 

\section{Introduction}

In astrophysical context, accretion is a process by which matter 
is collected around a central object.  Normally, when one talks about
accretion in binary systems, one star is tidally deformed 
and matter flows out from it to the compact companion. When one deals
with an isolated object, it may accrete from the interstellar 
medium at a very low rate. In many of the galactic centers
there are evidence of supermassive black holes. There are no companions,
but matter is accreted from winds of surrounding stars. In these
cases, stars may also be tidally disrupted if they come very close to the
black hole and the matter would be accreted from the disrupted 
star to the central black hole.

In the present review, we give an overall development of the subject
which took place over the last fifty years. It is clearly a Herculean
task and it would be impossible to cover the whole subject in these
few pages. We will first consider significant developments in last 
five decades and at the end discuss in detail about the
very modern developments.

Though the subject began with the study of accretion onto ordinary 
stars, very quickly  the similar method  was found to be useful for the
study of accretion onto compact objects, such as white dwarfs,
neutron stars and black holes. We mainly review the accretion flows
on black holes. Since winds are produced from compact stars
and accretion disks around black holes, and interestingly, are studied
from the same set of equations, the development of the subject of 
winds and outflows  began almost at the same time. So it is inevitable 
that some aspects of the outflows would be discussed simultaneously.

\section{Accretion and outflows problem over the decades}

\subsection{1950s: Toddler Days - era of curiosities}

The simplest problem of axisymmetric particle accretion through a 
shock front was done by Hoyle and Lyttleton back in 1939 [1]. 
The problem was to study how much matter would accrete on a 
star moving through an interstellar medium. The work
was not satisfactory as pressure effects were ignored. In 1952,
the classic paper of Bondi [2] was published. There he computed the
mass accretion rate on a star `in rest in an infinite cloud of gas'
by including the pressure effects. 
The solution showed clearly that matter, originating
subsonically at a large distance can become supersonic on a star surface.
Bondi [2] already anticipated some directions in  which the subject should
proceed: first, compressional heat may be lost and pressure at the inner edge
may be diminished causing much larger inflow rate. Due to this
non-adiabaticity, the polytropic index $\gamma$ should become
close to unity near the star. Second, a correct understanding may 
require a time-dependent solution.

Almost simultaneously, Biermann [3] suggested that the behaviour 
of cometary tails could be explained by assuming outflows from 
the solar surface. This was later explained by Parker [4] who showed
that one of the sub-sonic branches crosses to 
become supersonic and in fact, could explain the
velocity of solar wind at the Earth's orbit. This thermal pressure 
driven wind solution was thus found to be important in the subject of 
solar and stellar winds, and indeed for astrophysical jets, in general.
Parker's solution was thus similar to the Bondi-flow solution except that
the flow originates at smaller radius and goes to infinity. Also,
initial models were discussed for isothermal gas. 

From the observational point of view, 1950s saw a large number of 
pioneering results, obviously with ground based 
instruments. Jennison and Das Gupta [5] first resolved
%R.C. Jennison and M.K. DG, 1952, Nature, 172,996
two radio blobs of Cyg A and thus double radio sources were
discovered. Baade and Minkowski [6] termed the train of optical knots
seen in M87 as `jets' and proclaimed that they must be ejected from
%Baade, W. and Minkowski, R. 1954, ApJ, 119, 215.
the nucleus of the galaxy. These paved the way to a completely new subject of cosmic
radio jets and the physics around black holes.  These
powerful radio emitting regions turned out to be indirect evidences of
massive black holes at the centers of active  galaxies. 
In the 90s, evidences for such activities around 
smaller mass black holes (so-called galactic microquasars) were also discovered.

\subsection{1960s: Childhood - era of dare devils}

Having powered by very satisfying accretion model of Bondi [2], 
workers became engaged in applying this model to explain various 
known observations of star-gas interactions.
With improved instruments in the post-war period
radio observations became very popular and several 
important discoveries were made.  
However, Solar wind observations were much easier to
do and thus more advancements were made in improving wind solution by additional
physics.

In 1963, first QSO 3C272 was discovered [7, 8]. Using 
radio occultation of 3C273 by the moon, Hazard, Mackay and Shimmins [7]
using Parkes 210 feet radio telescope
found two distinct radio sources and located the position. This 
was communicated to M. Schmidt at Caltech 
who obtained the redshift by optical measurements [8]. This showed that the need to exchange data 
was already appreciated at this early age. In the first Texas Symposium of 1963 [9]
%Robinson et al, 1965
various explanations were proposed all converging to the idea that
the enormous energy radiated by a QSO must be gravitational in nature.
In the following year, Salpeter [10] used Schwarzschild 
%Salpeter, E. Apj, 1964, 140, 796 
solution of 1916 and interpreted that the luminosity could be due to Bondi 
accretion on very massive compact objects ($>10^7M_\odot$). These objects 
were termed as `collapsed matter' or collapsars at that time since 
the phrase `black hole' was coined by J. Wheeler only in 1967  [11].
The computed luminosity was found to be on the order of $10^{47}$ ergs per sec 
for reasonable parameters. The problem of local vs. cosmological nature of 
QSOs persisted, however. Hoyle and Burbidge [12]
debated over the possibility of cosmological distance since it was
difficult to explain rapid variation (40 percent in two years) of intensity
at 8GHz emission from 3C273.  

Russian school primarily headed by Zeldovich and Sunyaev had contributed immensely
in our understanding of the accretion processes. Indeed even today the 
so-called standard model is originated from this school. 
In 1966, Zeldovich and Guseynov [13] 
pointed out that there are several binary systems in which the 
companion is `unseen' yet massive indicating that 
collapsed stars (later known as black holes) should exist in these
systems. `An unambiguous proof of the existence of a collapsed star would naturally be of 
the great interest.' write  Zeldovich and Guseynov [13]. 
Today the situation has improved (see, [14]) and we have several very good
galactic and extragalactic black hole candidates.
Because Quasars have typical luminosities of $10^{44-47}$ ergs sec$^{-1}$, 
(which means destruction of $\sim 0.002-2$M$_\odot$ yr$^{-1}$ completely at each Quasar)
energy must be coming from gravitational energy.
Based on this, Lynden-Bell [15] made a strong argument that accretion close to Eddington
rate on super-massive black holes must be responsible for this tremendous activity. 

In 1963, a major discovery on the theoretical front was the vacuum solution
of Einstein equation around a compact rotating collapsed matter by R.P. 
Kerr [16]. Immediately after that there was an explosion of activities
in Cambridge, Princeton and other places to understand the implication
of this new solution. A large number of papers were published
to study behaviour of matter (mostly particle) trajectories (see, [17]).
Today, these objects are called Kerr black holes and 
are almost universally thought to be playing central role
in black hole astrophysics.
%R.P. Kerr, 1963, Phys. Rev Letters, 11, 522.

As observations were carried out, it became clear that QSOs 
are just special cases of Active Galactic Nuclei which come in 
a variety of forms: line-less BL Lacs, fainter Seyfert galaxies, Blazers,
Quasars, Optically Violent Variables, Liners etc. In the 90s
active efforts were made to unify these various classes.

Among other major discoveries in the 60s which are related to 
accretion/winds are the pulsars and X-ray sources. Antony Hewish 
discovered regular pulses of Radio waves towards the end of
1967 [18]. Among interesting models 
%[Hewish, A., Bell, S.J., Pilkinson, J.D.H., Scott, P.F., 
%Collins R.A., 1968, Nature, 217 709]. 
which were proposed immediately were emission from hot spots
on rotating compact stars by Ostriker [19] % (1968, Nature, 217, 1222) 
and gravitational lensing model (binary neutron stars with 
one occulting and amplifying other's radiation) of Barnothy
and Barnothy [20].

Unlike radio observations which were carried out in the forties and
in the fifties, X-rays from accretion onto compact objects could not 
be studied from ground. After the rocket technology was improved,
experiments started with Aerobee rockets. In 1962,
Giacconi and his team discovered X-Ray sources by rocket measurements.
During 1962-1966, they concentrated on Cygnus region, 
Sco X-1 and Crab Nebula, and determined their
positions.  Cyg X-1, the first black hole candidate 
was discovered in these early attempts in 1962 [21].
%(Giacconi et al, 1962)
%[Giacconi, R., Gursky, H., Paolini, F. and Rossi, B. 1962,
%Phys. Rev.lets, 9,439;
%Giacconi, R., Gorsky, H. and Waters, J.R. 1965, Natrere,
%207, 572] Gursky,H. Giacconi, R., Gorenstein, P., Waters, J.R.,
%E.J. Oda, M. Bradt, H. Garmire, G. Sreekantan, B.V. ApJ, 144, 1249, 1966]
From simple binary accretion model, soon it was realized that the 
companion should exist and optical search was also made in Palomar
plates. These were the first of  the experiments and the avenues
opened by these pioneers are still followed though the technology 
has advanced radically to observe much fainter sources for much 
longer periods using balloon and satellite technology.

Apart from professional astrophysicists and astronomers who are
interested in accretion physics, many others got interested in the 
black hole accretion as well. For instance, Salpeter [10] 
acknowledges discussions with R. Feynmann, M. Schwarzschild
and L. Spitzer. Even energy extraction by feeding rotating black holes
with negative energy through accreting matter
was evoked by Penrose [22] %[1969, Rev. del. Nuovo Cimento, 1, 252] 
in this decade. The prospect of Quasars being cosmological
became brighter each day leaving cosmologists such as George Gamow 
to wonder [9]: Twinkle, twinkle quasi-star/Biggest puzzle from 
afar/How unlike the other ones/Brighter than a billion suns/Twinkle,
twinkle quasi-star/How I wonder what you are.
Thus all the best brains of the day were working to sort  
out puzzle uncovered by the great discoveries of the 60s.

By early 1960s, the phrase `solar wind' became common and 
works were carried out to study interaction between solar
wind and the earth and comets.  The proximity of the Sun drove 
many astrophysicists to understand wind mechanisms better. 
Parker [23]
%(1963, New York, Interscience, Interplanetary Dynamical Processes; ) 
improved on the study of his earlier work [4] on solar and stellar winds.  
Important ingredients such as  angular motion and magnetic field of the sun
was added at this stage by Pneuman [24], % [?, 1966ApJ...145..242P.pdf], 
Weber and Devis [25] etc.
%Weber, E.J. and Devis, L. D. Jr. [v 148, 217, ***] etc. 
Weber and Devis [25] worked out the effects of angular momentum 
and radial-azimuthal field lines on the equatorial plane 
and computed the torque exerted on the sun and how the 
sun can be slowed down. Work advanced a great deal
in this direction with heating, cooling, conduction etc. For 
instance, Eisler [26] %[? Thomas, J.; 1969SoPh....7...46E.pdf] 
includes conductive heat flux in spherical geometry. Weber [27]
% [?, 1969SoPh....7..470W.pdf] 
discussed solution with conducting heat flux in flows with angular 
momentum and magnetic field. Whang, Liu and Chang [28]
% [?, 1966ApJ...145..255W.pdf] 
included viscosity and thermal conduction. Interestingly, they 
also discuss the ratio of the viscous stress to thermal pressure,
later known to be Shakura-Sunyaev viscosity parameter [29] in 
the context of accretion disks. In a classic paper by Axford 
and  Newman [30] solved the Bondi accretion 
flow and winds simultaneously after including viscosity 
and thermal conduction. They studies weak shocks as well.
More complex interaction which included bow-shock waves were 
also studied in this decade  [31]. Works with this vigour were not 
carried out in accretion  around black holes for another decade or so.
%[L. Biermann, B. Brosowski 
%and H.U. Schmidt; ? 1967SoPh....1..254B.pdf]. 

\subsection{1970s: Adolescence - Era of Confusion and Exploration }

In the 1960s, Pandora's box was opened on all fronts: Radio and
X-ray observations, Theory of black holes, even simple theoretical models 
of accretion flows. Models proposed were diverse: 
simplistic yet bold. The confusion persisted in the 1970s 
since the technology was mature to make satellite observations
and competition to build realistic models was too much. Hectic
search for a quick-fix was thus a necessity.

Models of accretion was going in several directions: 
the spherical accretion was simple and had a plenty 
of room for improvements. Detailed spectra of stars and QSOs
demanded that angular momentum should be included and mechanism 
of radiation from the disks should be more efficient than 
that of a spherical accretion. Thus came the standard 
Keplerian disks of Shakura and Sunyaev [29]. Cosmic radio
jets were observed extensively and along with it came the 
realization that there must be powerful engines sitting at the
galactic centers to produce them. Since black holes have
no hard surfaces, jets must be produced from the accretion itself. 
Disk models evolved specifically to produce jets efficiently 
and thick accretion disk models began which were perfected 
and studied in detail in next decade. Below we discuss 
briefly these models. From the work of Lynden-Bell [15]  it was 
realized that the QSO emission must be related to 
accretion onto black holes. In 1970, from the study of 
absorption lines, Sturrock [32] suggested that the QSO 
themselves must have formed these jets and outflows.
So, naturally there was an extra emphasis to understand the
origin of large luminosity and the origin of jets simultaneously.
This has only recently been understood in full detail 
which will be discussed later.

It is to be noted that though black holes themselves
were new at this time, numerous authors ({\it e.g.} [33]) % Wyller, 
already were engaged in looking for possibilities 
to detect them from Doppler shifts of line emissions. 
So, many of the ingredients of what to look for (such that
double-horn pattern of line-emission found in black 
hole accretion flow) were already emerging. 
Bardeen [34] % (JM) [1970, Nature, 226, 64 ] 
suggested that by the time the mass of the accreted matter is 
more than fifty percent of the  initially non-rotating 
black hole mass, the black hole would likely to become
extremal Kerr black hole ($a\sim 1$) though radiation effects
could slow it down to more like $a=0.998$ [35]. % (Thorne, 1974, 1974ApJ...191..507T).
A treatise on general relativity and particle and light
behaviour around black holes which was compiled by Misner, Thorne and Wheeler 
in 1970, published in 1973 [36], even tried to convince reader through a dialogue
between Sagredus and Salvatius why the phrase `black hole' is the most
appropriate for the collapsed stars.

Pringle and Rees  [38] % [1972, 1972A+A....21....1P] 
first attempted quantitative analysis of the
nature of the X-ray radiation which are 
emerged from an accretion disk around a black hole or a neutron star. 
Even the nature of accretion on a neutron star
with magnetic field and origin of pulsed emissions were discussed. They conjectured
that both for high and low accretion rates, the inner part of the disk should
flare up because of high radiation pressure and because of low efficiency of 
radiation respectively. This work is clearly precursor of radiation pressure
dominated thick disks and ion pressure dominated accretion tori respectively.
In the same year, in 1972, Shakura and Sunyaev [29] % [1973, 1973A+A....24..337]
wrote the classic paper on the standard disk model, in which they assumed that the angular 
momentum distribution is Keplerian  throughout the disk. For stability reason, 
the inner edge of the disk was truncated at $6GM/c^2$, i.e., $3$ Schwarzschild radii. 
Viscosity was assumed to be responsible to transport angular momentum outwards to give rise
to such a distribution. The heat generated by viscosity was assumed to be radiated away
instantaneously from the surface. The viscous stress was assumed to be dominated by
the $w_{r\phi}$ component (which is a reasonable assumption for a geometrically thin
disk) and whether it is turbulence viscosity or magnetic viscosity or both,
the viscous stress is considered to be proportional to the 
local pressure by the relation $w_{r\phi}=-\alpha \rho v_s^2$, where
$\alpha$ is a constant, $\rho$ is density and $v_s$ is the isothermal sound
velocity. For causality, $\alpha <1$ always. The disk was assumed to be 
in vertical equilibrium and the vertical thickness was computed accordingly.
In presence of strong radial motion $v_r$, Chakrabarti and Molteni [38]
suggested that the above prescription should be modified to
$w_{r\phi}=-\alpha \rho (v_s^2+ v_r^2)$. This would 
create a smooth variation of the viscous stress across 
a shock transition and will not transport angular momentum 
unnecessarily by an axisymmetric shock.

Shakura and Sunyaev [29] presented analytical solutions of the disk variables such as
temperature, density etc. as a function of the radial distance from the
axis. They also computed radiation spectrum by adding black body 
contribution from successive annulus of the disk. This so-called multicolour
black body spectrum became the hall mark of the standard disk model.
The above work was carried out using Newtonian equations. Novikov and
Thorne [39], subsequently extended this  work to include general relativity.
Along with the continuum emission, workers began to observe line emissions 
from a rotating accretion disk [40] % [1974, Meszaros, P, 1974A+A....35..171]
around a black hole. The Doppler broadening, Doppler shift and gravitational 
broadening were also studied in this work.
At roughly the same time another study envisaged outflow driven by
X-ray radiation on the disk [41]. % [D.T. Wickramasinghe, 1974MNRAS 168 297].

Meanwhile in  December 1970, the UHURU satellite was launched. This was the 
first Astrophysics related satellite and first few things which were
looked at included the black hole candidate Cyg X-1. First few 
papers [42] % [1971, R. Giaccono, E. Kellogg, P. Gorenstein, H. Gursky, and H. Tahanbaum, APJL 165, 27] 
showed that X-rays of energy larger than 100keV was emitted by this source. 
There were contradictory reports on whether this source 
showed periodic variability  with $73$ms variability or not [43].
%[E. Schreier, H. Gursky, E. Kellogg, H. Tananaum, R. Giacconi, APJL, 170, 21].
Only recently, confirmed  very low-frequency (0.05-0.1Hz) Quasi-Periodic oscillation
has been reported from this candidate [44]. %(1994ApJ, 424, 395, Vikhlinin et al.].

Soon after the standard disk model was  published, Lightman and Eardley [45]
%[1974, 1974ApJ...187L...1L.pdf] 
pointed out that in a radiation pressure dominated
region, viscosity prescription  of Shakura and Sunyaev was inconsistent if
$\alpha$ remains constant. In fact, they show that the disk would break into thin rings. 
Subsequently, Thorne and Price [46] %(1975, ApJL, 195, L101) 
argued that Shakura-Sunyaev 1973 disk model is unable to explain the 
high energy X-rays found in Cyg X-1. Eardley, Lightman and Shapiro [47] % (1975, L153, 199, ApJL)
had a very successful model with a Keplerian disk which flares into
a thick disk closer to the inner edge. They considered two temperature 
flow and it was a single component disk with a small accretion rate. 
The spectrum of Cyg X-1 was fitted reasonably accurately. Later, Pringle [48]
%(J. 1976MNRAS.177...65P.pdf) 
showed that such a disk is thermally unstable.
Observed spectra of the black hole candidates, such as Cyg X-1 [49]
%(Sunyaev and Tr\"umper, 1979)
indicated that the spectrum consists of two distinct components. 
Observations similar to Cyg X-1 were made for active galaxies and Quasars [50].
%(Katz, 1976; ApJ, 206, 910). 
The soft X-ray bump in these spectra could be explained by multicolour black
body emission from a Keplerian disk. The power-law component
of the spectra was explained by Comptonization of softer photons by
hot electrons from a magnetic Corona on an accretion disk [51].
%(Galeev, Vaiana and Rosner 197**).
Not only does the spectrum consist of two distinct components, 
it was apparent from X-Ray observations [52]
%(P.C.Agrawal et al. 1972 Ap. Space Sci, 18, 408; Tananbaum, H. et al. 1972, ApJ Lett, 177, L5)
that Cyg X-1 actually has a `soft state' (when X-ray power is emitted in soft 
X-rays) and a `hard state' (when X-ray power is emitted in hard X-rays).
Ichimaru [53] %(S.; 1977, ApJ, 214, 840) 
pointed out that the two states observed in
Cyg X-1 could be due to two distinct solutions that the equations yield --
one is an optically thin branch which emits hard X-rays and the other is an
optically thick branch which emits soft X-rays. The work was done
using Newtonian equations and thus the inner boundary conditions were not
appropriately chosen. The optically thin solution
was later termed as Advection Dominated Flow or ADAF [54]. % (Narayan and Yi, 1994).
Hoshi and Inoue [55] % (R. Hoshi, H. Inoue, PASJ, 1988, 40,42) 
considered X-ray irradiated disk and also 
found similar transition from optically thin to optically thick
solution. The bimodal behaviour was claimed to be responsible for the
state transition. Today, it is understood that the hard and soft components of the 
spectrum are emitted  at the same time and can vary
independently and these problems are easily explained by 
two-component advective flow (TCAF) model discussed later [56] in this review.

Progress in general relativistic accretion disks occur
greatly by the efforts of Bardeen and Patterson [57] % [1975ApJ...195L..65B] 
who studied disk accretion in an inclined plane at a large distance and showed that 
the flow is dragged in the equatorial plane of the black hole in the region between $100GM/c^2$
and $10^4 GM/c^2$. This kind of tilted disk was envisaged  sometime back 
by Wilson [58] %[1974, J., ApJ 187, 575] 
who suggested that during asymmetric explosion of a supernovae, resulting compact 
object may change its spin axis from that of the normal on the binary plane. 
The effect of Kerr geometry on the accretion disks was explored in great detail 
in works of Cunningham [59]. % (CT, 1975, ApJ, 202, 788). 
For instance, how the spectrum is modified due to gravitational red-shift
was investigated. Another curious work is that of Luminet [60] % (1979, JP, AandA, 75, 228) 
who computed how an accretion disk around a black hole itself would look like
if photographed from the vicinity. Due to the light bending effect, the disk would have 
an interesting shape --- the region away from the observer would be bent and even 
lower side of the disk could be seen. Matter on
one side of the disk moving away from 
the observers would emit the red-shifted light, and that from 
the other side would emit  the blue-shifted light. 

Along with the accretion disk model, efforts were being made to compute 
the properties of spherical symmetric Bondi accretion in presence of heating and 
cooling effects. Michel [61] % [1972, F.C. Michel, Ap and Sp. Science, 15, 153] 
extended the original adiabatic work in general relativity. Shapiro [62] % [1973a, 1973ApJ...180..531S,] 
included electron-electron and electron-proton bremsstrahlung effects. 
Using the suggestion of Shvartsman [63] % [1971, V.F. Sh.., 1971, Sov. Astr. -AJ, 15, 377] 
and Zeldovich and Novikov  [64] % [1971, Ya. B / I.D., Relativistic Astrophysics,
Shapiro [65] % [1973b, 1973ApJ...185...69] 
included effects of tangled magnetic field whose energy density could be as high as
gravitational energy density and found that the synchrotron radiation will be
predominantly emitted in the infra-red region. An important addition to such 
study is the effect of pre-heating on the luminosity  [66]
%[Ostriker, JP, R. McCray, R. Weaver, A. Yahil; 1976ApJ...208L..61O.pdf] 
of the spherical emission.
They find that if the infalling gas intercepts X-rays emitted at the inner 
edge of the disk, then it will be heated up if the sound speed exceeds the
escape velocity, the gas accretion is decreased resulting in decrease in the 
luminosity limit  orders of magnitude lower compared to the Eddington limit. 
If the efficiency $\eta << 1.7 \times 10^{-4}$, the results agree with those of Shapiro [62],
but as $\eta \rightarrow 1.7 \times 10^{-4}$ their limit is higher compared to 
Shapiro [62] who ignored the effects of pre-heating. Later,
Bisnovatyi-Kogan and Blinnikov [67] % (G.S., S.I., 1980, MNRAS, 191, 711)
repeated this calculation and found that solution exists for 
any efficiency of energy release, including those 
typical for accretion on neutron stars, for any luminosity
below the Eddington limit. These conclusions [66, 67]
%Ostriker et al.  and Bisnovatyi-Kogan and Blinnikov
were questioned by Stellingwerf and Buff [68] %(R.F.; J.; ApJ, 260, 1982, 755) 
who found a much smaller forbidden region and showed that the flow is
disrupted only if the unheated Bondi luminosity 
exceeds the Eddington Luminosity. Spherical Bondi flow
was also studied on neutron stars as the procedure was the same, but the 
inner boundary condition varied. The boundary layer of a neutron star where the
flow settles down to zero velocity was first studied by Shapiro and Salpeter [69].
%[1975, 1975ApJ...198..671S.pdf]. 
Thus the importance of the 
inner boundary condition was recognized, and particularly it was realized that
the boundary layer is an extension of the disk flow itself. Though for
black holes there are no hard surfaces, this concept could be extended [70]
%(Chakrabarti et al, 1996 PR) 
for black holes also, especially if matter possesses angular momentum,
albeit small. This will be discussed later.

Yet another type of study included giving a pre-assigned velocity and density
profiles and obtaining the corresponding variation in temperature as a function of
radial distance in presence of bremsstrahlung, synchrotron etc. [71].
% [Meszaros, 1975, 1975A+A....44...59M.pdf]. 
This type of work was extended to two temperature flows later by Maraschi, Roasio and Treves [72].
%[1982, L, R, A, ApJ, 253, 312].
In the absence of a fully self-consistent study, these solutions 
gave pretty good idea about how the electron temperature gradually deviated from the
proton temperature due to faster cooling processes. Eardley and Lightman [73]
%[D.M. Eardley and A.P. Lightman 1975 {\it Astrophys. J.}  {\bf 200} 187]
subsequently solved two temperature Keplerian disk.
Till today, there are debates on  whether electrons and protons should have 
separate temperatures or not, particularly that there are processes
(Ohmic heating) which may heat up the electrons otherwise cooled
down by synchrotron emission  [74].
%(Bisnovatyi-Kogan, GS; Lovelace, RVE, ApJ, 1997, 486, L43).

With the understanding that general relativity should be essential to explain
behaviour of matter and radiation around a black hole both the accretion and
wind models were generalized, first by assuming 
predominantly rotating flow [39] %(Novikov and Thorne, 1973) 
and then by assuming predominantly radial flow [75].
%(G.R. Blumenthal and W.G. Mathews, 1976ApJ...203..714B.pdf). 
The later work was essentially the generalization of the Bondi flow and Parker winds
already described. Moreover, the shock transitions as described by 
Holzer and Axford [76] in the context of solar winds were also studied here
% (T.E., W.I.,1970, ARAA, 8, 31) 
in the context of accretion. A supersonic flow must have a shock transition 
if the fluid pressure in the upstream has a finite 
value. Thus accretion on a neutron star, or winds propagating supersonically to a
medium of finite pressure must have a shock. In the case of black hole accretion,
the pressure is essentially zero on the horizon, and hence a spherically 
accreting adiabatic flow endowed with a single sonic point does not produce a shock
in accretion.  

The decade of the 70s saw another set of studies which were essentially
computer simulations. As new and faster computers came in 
laboratories, one started carrying out {\it numerical experiments} by computers. 
This is because it is understood that it is impossible 
to perform real experiments in Astrophysics.
Time dependent general relativistic flows  were studied in this
decade by pioneers such as Wilson [77]. % (JR, 1972, ApJ, 173, 431). 
Wilson used his skill of simulating collapse of stars and supernovae explosion and applied them to
study accretion of rotating matter around a Kerr black hole. He found the
existence of propagating shocks inside the disk when angular momentum is 
significant. 
He kept the angular momentum to be constant, contrary to standard works 
[29, 39] which immediately followed his work.
Modification of the equations of Shakura and Sunyaev [29] clearly included
addition of the pressure terms and advection terms and more realistic heating
and cooling effects. Maraschi, Reina and Treves [78]
%(L., C., A., 1976ApJ, 206, 295)  
included radiation pressure and drew a few important conclusions. They found
that as accretion rate is increased even higher than the Eddington rate, the 
accretion solution still exists. Furthermore, with the radiation pressure,
the flow is deviated from 
a Keplerian flow. This work was, however, carried out using Newtonian geometry. 
Lynden-Bell [79] %(D. 1978, 17, 185)
worked out the steady state structure of the flow
with constant angular momentum in general relativistic flows 
and predicted that the flow would develop giant vortices along the axis,
simply because centrifugal force would keep matter away
from the axis. Here pressure terms were included as well (which was
unimportant in Keplerian disks). 

While this decade was full of theoretical advancements, one should not forget that
space based experiments were already in place and major results were arriving
almost on a daily basis. Similarly Radio telescopes were in place since 1940s and 50s
which were being pointed at the Quasars and active galaxies. Cosmic radio
jets, which were found to be ejected from galactic centers were highly puzzling
and it was already realized that they must be related to the central engines of 
the active galaxies and specifically to the accretion disks. 
First set of serious models  of cosmic jet formation were given by 
Blandford and Rees [80] % (1974, MNRAS, 1974, 169, 395)  
and Scheuer [81]. % (1974, PAG, MNRAS, 176, 421).
Blandford and Rees [80] envisaged that somehow matter would bore `de-Laval' type
nozzles through ambient medium and propagate supersonically on either side  of the galactic plane. 
Observations of these jets put pressure on theorists to modify their models
so that super-luminal and well collimated jets could be produced.

Meanwhile, it was realized by several workers [46, 47, 82]
%(Thorne and Price, 1975, Eardley, Lightman and Shapiro, 1975, Stoeger, W.R, 1978 MNRAS 182, 647S) 
that radiation pressure may cause disks to thicken due to thermal and other instabilities
and an era of thick accretion disks followed.
Key issues which motivated to construct this new disk models 
were: (a) pressure, either due to radiation or due to hot ion,
must be included which was ignored in the standard Keplerian disk 
models of Shakura and Sunyaev [29] (b) matter must accrete very rapidly towards a 
black hole and thus the infall velocity should be included and finally
(c) formation and collimation of cosmic jets must be explained. 
Lynden-Bell [79] simply assumed constant angular momentum flow
and found that disks thicken up closer to the black hole.
This work was extended  and thick accretion disk models were 
constructed by Paczy\'nski and his collaborators [83-85] by choosing various
%Kozlowski, M., M. Jaroszynski, and M. A. Abramowicz,
%1978; M. Abramowicz, M. Jaroszynski and M. Sikora, 1978; Paczynski and Wiita, 1980). 
combinations of angular momentum distribution. 
In these so-called radiation pressure supported `thick-disk' models,
radial velocity was ignored  totally. These disks had `vortices' or `funnels'
which were hoped to be suitable for launching jets and outflows.
These disk models were not self-consistent, in that they had an element of ad-hoc
angular momentum  distribution, but their importance lie in that they were pre-cursors of more 
complete models which include both rotation and advection. These are discussed later.

\subsection{1980s: Youthful days - era of enlightenment}

In the 70s, a large number of theoretical fronts were opened. In the 80s most of these
fronts were consolidated and bits and pieces were put together to have a more global view,
which is, after all, the goal of all the model builders.  
Using simple polytropic equation of state, the spherical accretion solution
in Schwarzschild geometry 
was solved by Begelman [86] who showed 
that there should be one critical point. This was farther
extended to relativistic equation of state by Brinkman [87]. %(W, 1980, AandA 85, 146-148). 
Moncrief [88] % (V, 1980, ApJ, 235, 1038) 
showed that the transonic spherical flow is stable as well. 
Works proceeded on non-adiabatic flows also. Pre-heating was already discussed [66].
%by Ostriker et al. (1976). 
Effects of dissipation heating by magnetic field
was studied by Scharleman [89] %(E.T.; ApJ, 1981, 246, L15)
who showed that sound speed can become so large that the Bondi flow would cease to be 
supersonic. Of course, this would lead to shock waves or non-steady flows.
Yahel and Brinkman [90] % (R.A.;W.; ApJ, 244, L7) 
studied the emergent spectrum from a spherical Bondi flow and also studied the effect of $e^--e^+$
on the emergent spectrum and found a sharp cut-off at around $1$MeV
due to $\gamma-\gamma$ interaction. Thorne and his collaborators [91]
%(K.S. Thorne, R. Flammang, A. Zytkow, 1981, MNRAS 1994, 475) 
presented detailed general relativistic photo-hydrodynamical
spherical accretion in optically thick regime. Flammang [92]
%(RA; MNRAS, 199, 833, 1982) 
studied the nature of critical points in presence of radiative transfer.
Ipser and Price [93] %(J;R; ApJ, 255, 654, 1982) 
considered the effect of synchrotron cooling in Schwarzschild geometry
and classified different regimes of accretion according to
whether synchrotron self-absorption is important or not.
Payne and Blandford [94] %(D.G.; R.1981 MNRAS, 196, 781)
studied the nature of the emitted spectrum of the optically thick converging flow
in Newtonian geometry and found that the energy spectral index $\alpha$ ($F \sim \nu^{-\alpha}$)
would be more like $2$. With general relativistic considerations this index becomes
$1.5$ and would turn out to be the most convincing signature of a black hole
accretion [56]. A more general solution,
though in Newtonian geometry, was put forward by  
Wandel, Yahil and Milgrom  [95] %(A;A;M; v. 282, p 53, 1984, ApJ)
and Colpi, Maraschi and Treves  [96] % (M;L;A; ApJ, 280, 319, 1984)
who studied single temperature solution of non-adiabatic flow and showed 
that Quasar luminosity increases
with the accretion rate and the luminosity is a few percentage of the Eddington luminosity.
A two temperature spherical accretion solution with bremsstrahlung, synchrotron cooling, 
Coulomb coupling between protons and electrons and and Comptonization was constructed. 
It was seen that the protons could be hotter than the electrons by a factor 
of a thousand or so. However the spectrum of Seyfert galaxies and Quasars
where clear excess of radiation in ultraviolet was seen [97] 
%(MA Malkan, Sergent, WAW; 1982, ApJ, 254, 22) 
could not be explained by spherical or
converging flows. Fortunately, Shakura-Sunyaev Keplerian disk [29]
is known to emit the multi-colour blackbody radiation in this region and
several cases were fitted and satisfactory results were found [98].
%(Sun WH, and Malkan MA, 1989, ApJ, 346, 68).
The mass of the black hole was found to be about $10^8M_\odot$.
Subsequently, Wandel and Petrosian [99] % (A: V; 1988, ApJ, 329, L11) 
showed that  
basically two parameters, namely the mass of the black hole and the accretion rate
can describe the spectrum very well and the masses of black holes of the Quasar
and Seyferts fall in the range of $10^8-10^{9.5} M_\odot$ and $10^{7.5}-10^{8.5}M_\odot$
respectively. Ross, Fabian and Mineshige [100]
%(RR, AC, S, MNRAS, 1992, 258, 189)
improved the disk model by properly treating the inner edge of the
accretion disk and showed that even for disks around a massive black hole,
a significant radiation could be in soft X-rays. 
Generally, active galactic nuclei also show line emissions along with 
continuum emissions. These lines are believed to be emitted from
rapidly moving clouds on both sides of an accretion disk. Measurement 
of the motion of the cloud from the Doppler shift and the distances 
of the cloud from reverberation mapping [101]
%(Blandford and McKee, 1982)
method can give an estimate of the mass of the central object. In this method,
the time lag between certain variation in the continuum spectra and the
line emission is used to measure the distance of the broad line emitters.
Masses of a few active galaxies have been measured this way:
NGC 5548 ($8.8 \times 10^7 M_\odot$), NGC 3227 ($3.8 \times 10^7 M_\odot$).

There was some degree of uneasiness from several fronts: Astrophysicists
having expertize in stellar accretion and  winds would have liked to understand 
the black hole accretion  as well, but not everyone was an expert in general relativity.
For instance, studies of oscillations of stars were extended to accretion disks
around white dwarfs by Cox [102]. % (1981, JP; ApJ, 247, 1070). 
This was done to explain
quasi-periodic oscillations (QPOs) of dwarf-novae seen before. The extension to this work
to relativistic disk was done much later by Wagoner and his collaborators  [103]
%(M A Nowak and RV Wagoner, ApJ, 1991, 378, 656)
and topic disko-seismology was thus introduced. Other interested individuals in the
oscillations and QPOs were Kato [104]
% (S.; MNRAS, 1978 185, 629) 
and Livio and Shaviv [105]. % (M.;G.; 1981, 244, 290). 
While Kato, Honma and Matsumoto [106] % (S; F; R;  Yamada meeting) 
later considered trapped oscillations as potential candidates for 
Quasi-Periodic oscillations, Langer, Chanmugam and Shaviv [107] %(S.H.; G.; G., ApJ, 258, 289, 1982)
considered shock oscillation in the accretion column
to be the serious candidate. It will be shown later
that QPOs in black hole oscillations are likely to be due to shock oscillations in accretion disks [108].
% (Molteni, Sponholz, and Chakrabarti, 1996).

Paczy\'nski and Wiita [85] %(1980) 
while studying accurate description of 
thick accretion disks  presented a potential $\phi=-GM/(r-2GM/c^2)$, where $G$ and $c$ are the
gravitational constant and velocity of light respectively, $M$ is the mass of the
central black hole. This potential has the feature that Keplerian distribution of 
angular momentum is exactly the same as that around a Schwarzschild black hole. Thus
it was expected that work done using Newtonian formalism with this
so-called pseudo-Newtonian potential might yield results similar to those 
obtained around a black hole. This has indeed been found to be the 
case since the error occurring is at the most a few percent [109] and perhaps
within the observational error bars. A similar potential was constructed by
Chakrabarti and Khanna [110] % (1992) 
which was much improved by Chakrabarti and Lu  [111] %(1993)
to work in Kerr geometry. A second problem was that 
workers already realized ({\it e.g.}, [112]) %W.R. Stoeger, 1980, 235, 216) 
that the disk must not remain Keplerian close to a black hole
as the radial velocity becomes very large below the marginally stable
orbit ($r=r_{ms}$). Both the approaches [85, 112]
%of Paczy\'nski and Wiita (1980) and
%Stoeger (1980) 
assumed that either the thick disk or the 
sub-Keplerian flow are generated from a Keplerian disk nearby. 
More work along this line, this time including  radial velocity
around the Paczy\'nski-Wiita potential, were carried out by
Paczy\'nski and his collaborators [113]
% (e.g., Paczy\'nski and Bisnovatyi-Kogan,
%1981, Acta Astronomica, 31, 1981; B. Muchotrzeb and B. Paczynski,
%Acta Astronomica, 32, 1982, 1). 
The global solutions were not found properly, but the indications were clear 
that the flow has to cross a sonic point close to a black hole. Similar study 
to find a consistent solution for $r<r_{ms}$ was also attempted by 
[114]. % Matsumoto et al. (1984).
Abramowicz et al. [115] % (1988) 
in the so-called `slim' accretion disk tried to introduce
transonicity in the global solution and was only partly successful and many 
un-physical solutions were also obtained. More complete and global solution
of the slim disks were obtained by Chen and Taam
[116] % (XM; R; 1993, ApJ, 412, 254)
who showed that a significant heat generated by viscous friction is 
carried away inside the black hole.

At the same time the radiation pressure supported disk models were being
explored, Rees and his collaborators [117] % (e.g., Rees, Begelman, Blandford and Phinney, 1982)
argued that for small accretion rates, the disk should be so inefficiently cooled
that the gas temperature could rise to almost virial temperature and the 
ion pressure could puff up the disk to form a so-called `ion-pressure supported' 
thick accretion disk. Since thick disks have a  toroidal `center' where 
the pressure is maximum, they behave like very hot stars, only toroidal
in shape. Compared to stars where the central temperature is around $10^7$K, disks
can have temperatures around $10^{8-11}$K, though densities are
much lower in disks than in stars. However, nuclear reaction rates
are very sensitive to temperatures and thus even if the 
infall time scale in disks are low (in the hot regions in stellar black holes,
this time is around a few milliseconds) the nuclear reaction could be significant
to modify the composition of the infalling matter.
Rees [118] first suggested that hot ion tori can have some nuclear reactions and
energy generation. Chakrabarti [119] generalized the 
thick accretion disk study in any axisymmetric spacetime and showed that 
the disks and the pre-jet matters could be studied using the same
theoretical formalism. The angular momentum distribution was
based on the properties of the von-Zeipel surfaces on which 
angular momentum and angular velocities remain constant. Using this model [120],
Chakrabarti  and collaborators [121]
%(1986), Chakrabarti, Jin and Arnett (1987) and Jin, Arnett and Chakrabarti (1989) 
computed the nuclear reactions inside thick disks in 
detail and showed that thick accretion disks with a very low viscosity could 
have significant nucleosynthesis and a part of the newly formed elements could 
be ejected by jets and are dispersed in the interstellar space creating 
some metalicities. These works were further followed up by several workers
and similar results were reported [121].
% (Arai and Hashimoto, 1992, 254, 191 AA; Hashimoto et al., Eriguchi and Miller). 
Recently, with the perfection of 
the transonic flow solution (namely, with the emergence of the two component 
advective flow [TCAF] paradigm [56]) this problem was taken up once more [122]
%(Mukhopadhyay and Chakrabarti, 2000) 
and it was observed that in presence of 
standing shocks the possibility of nucleosynthesis is much more. It was also
observed that the flow indeed remains stable in most of the parameter space
even when nuclear reaction is turned on.

Since the properties of thick accretion disks are closely linked to 
the deviation of the angular momentum distribution from the Keplerian 
distribution, some curiosities arose as to whether self-gravity of the
thick disk itself would cause any further deviation. After all,
the disk has a center (of toroidal topology) and matter between the
black hole and the center could require lesser angular momentum to stay in
stable orbits, while matter outside that of the centre would require
more angular momentum than Keplerian due to combined gravity of the
black hole and the disk.  This was first analyzed in Newtonian geometry [123].
%by Abramowicz et al. (!984, MAA, Anna Curir, A, Schwazenberg-Czerney, R.E. Wilson;
%MNRAS, 1984, 208, 279). 
In general relativity exact solution of the space time metric
is possible in presence of an axisymmetric ring [124]
%(Chakrabarti, 1989, {\it Journal of Astrophysics Astronomy}, {\bf 10}, 261) 
and several works in this connection were carried out later [125].
%(Y.; E; Eriguchi and Mueller,  1993, ApJ, 416, 666). 
 
Just as the progress in thick disk was being made,
a  parallel line of approach was adopted which abandoned the concept of a Keplerian disk 
altogether and consider a self-consistent solution with a simple angular momentum
distribution. Liang and Thompson [126] using the Paczy\'nski-Wiita potential argued that unlike
a Bondi flow, a thin, rotating, adiabatic flow can have three sonic points, two are of `Saddle' type and
one is of `Center' type. From this, one notices
that in order that such an adiabatic flow passes through a sonic point the flow must have angular
momentum less than a Keplerian flow [127]. However, it was wrongly concluded that 
because there were two saddle points, the flow would not know which one to use, and
possibly exhibit bi-stability behaviour [127]. This confusion was removed later by Chakrabarti [128]
where it was shown that the entropy densities at these two saddle points
are completely different and such bi-stabilities should not occur. 
Possession of more than one saddle type sonic point prompted the study of shock 
waves in the disks just as there were studied in solar winds [76].
%(Holzer and Axford T.E., W.I.,1970, ARAA, 8, 31).
Already Chang and Ostriker [129] pointed out that the presence of pre-heating 
in a spherical flow would change the speed of sound in the flow, causing 
the flow to have more than one sonic point
giving rise to standing shocks. These shocks were very weak
and located very far out. Nevertheless, they increased the efficiency of emission
in spherical flows. Fukue [130] pointed out that indeed shock solution in rotating  accretion flows is
a possibility. Chakrabarti [128] studied in detail such flows and found that both
accretion and winds can have shocks in a large region of the parameter space
spanned by specific energy and angular momentum. He separated the parameter space according to 
whether shocks could form or not. One of the most interesting aspects 
of these solutions is that perfectly stable and global solutions
seem to be existing even without viscosity. This was because for a black hole
accretion, the gravity eventually wins and allows matter to sink in
even with constant angular momentum. Another important point is that
like a Bondi-flow, the energy need not be dissipated at all along the way to a black hole.
The entire amount of energy which the flow possesses at infinity can 
enter inside a black hole without radiating anything! This aspect is very important
since the event horizon sucks in matter and energy and a large accretion rate
not necessarily translates into luminosity. This is in contrast with solutions
around neutron stars (or any other star for that matter) where half of the
kinetic energy is released  on the boundary layer. Recent observations [131]
suggest that black hole candidates indeed show very low luminosity 
for a given accretion rate (particularly for a low accretion rate
for which intrinsic cooling is low) and therefore testifying the correctness of the
advective flow solutions of Chakrabarti [128] and its more generalized version
developed by him in the 1990s. In any case, formation and study of shock waves in 
accretion disks turned out to be very important in black hole astrophysics 
and we plan to discuss about them later when we describe achievements in the 1990s.

Progress in the direction of numerical simulation of accretion disks were also very good 
in the 80s. Pringle [132] demonstrated how viscosity
can transport angular momentum and allow matter to sink
inside a Keplerian disk. Vitello [133] % (P; ApJ, 1984, ApJ, 284, 394) 
considered time dependent simulation of spherical flows and found the solution to 
quickly converge to the steady state flow. The net escaping luminosity 
was found to be very small compared to the Eddington luminosity.
We already pointed out that as far back as 1972, Wilson 
considered the simulation of {\it rotating} adiabatic disk-like flows and found propagating 
shocks. With the improved numerical skills and computational
power, Hawley, Smarr and Wilson [134] considered the same problem and
found very strong accretion shocks forming near the centrifugal barrier.
However, the codes were dissipative and  parameters in which shocks could form 
were not known until the work of Chakrabarti [128]. As a result,
in all the numerical works these shocks were found to propagate backward. Also, the theoretical 
was not known, and therefore the only available solutions to compare the numerical results
were those of thick accretion disks which had no radial motion and no standing shocks.
Thus, the comparisons were also poor. Another set of simulations were carried out
with radiative flows by Eggum, Coroniti and Katz [135]. They carried out 
simulations for both the sub- and super-Eddington accretions. For sub-critical Keplerian accretion
disks they find that initially radiation pressure supported disks
quickly become gas pressure supported and collapsed on the equatorial plane, as expected from 
works of Lightman and Eardley [45]. A super-critically accreting flow, on the contrary,
was found to form a thick disk with outflows along the vertical axis. 

Although we focussed our discussion on numerical simulations on the
disk-type accretion flows, a significant number of papers 
have been written on axisymmetric flows which form because of 
motion of a star through a medium. The original problem goes back to 
Hoyle and Lyttleton [1]. Numerical simulation
study of such axisymmetric, but non-rotating, accretion was carried out by
Matsuda et al. [136] % (T. Matsuda, M. Inoue and K. Sawada, MNRAS, 226, 1987) 
and Taam and Fryxell [137]. % (R, B., ApJ, 331, L117, 1988). 
The solution was seen to be unsteady and accretion seemed to be occurring with a `flip-flop' instability.
These simulations were carried out in two dimensions, but later work in the 90s in three
dimensions and higher resolutions do not show these instabilities very much [138]. 
%( Matsuda, T., Ishii, T. Sekino, N.  Sawada, K., Shima, E., Livio, M. and Anzer, U.; 1992, 255, 183;
%M. Ruffert, ApJ, 427, 1994, 342; M. Ruffert and Arnett ApJ 427, 1994, 351).

Theoretical results of the 70s were quantified and actual line emission
profiles of a relativistically rotating moving accretion disk
around black hole was calculated by Garbal and Pelat [139]. % (D.; D; AA, 95, 18, 1981)
It was shown that each emission line should be split into two lines,
one blue-shifted and the other red-shifted, the blue shifted one being
more intense due to Doppler boosting. Similar calculations
were carried out by Smak [140] %(J., Acta Astronomica, 1981, 31, 25)
and elaborated and extended later by Horne and his collaborators [141]. 
% (K.; T.; MNRAS 218, 1986; Marsh and Horne, 1988, MNRAS 235, 269). 
Doppler Tomography technique introduced by these authors is 
able to determine the velocity profile on the accretion disk very accurately,
and is a good tool to verify if the motion is strictly Keplerian or not.
In some of the broad line radio galaxies, such as 3C390.3 and ARP102B [142]
%(e.g., K. Chen, J.P. Helpern, ApJ, 344, 115) 
the double peaks were found to vary with time in a manner that is perhaps
inconsistent with a relativistic Keplerian disk and may require
spiral shocks [143]. % (Chakrabarti and Wiita, ApJ, 434, 518, 1994). 
Similarly, in the ionized disk of M87, spiral patterns 
were observed  [144] %(Ford et al. ApJ, 1994, 435, L28) 
which could also be modeled with non-Keplerian disk [145]. % (Chakrabarti, ApJ, 1995, 441, 576)

As in the case of thin accretion disks in the 70s, thick accretion disks 
were also found to be unstable under lower mode non-axisymmetric 
perturbation due to strong shear [146],
%(Papaloizou and Pringle, 1984, JCB; J.; 208, 721), 
especially when the angular
momentum is constant or nearly constant. Numerical simulations
devoid of radial motion did show formation of four planet-like
structures due to balance of pressure gradient force and 
Coriolis force in such accretion disks [147].
% (Hawley, JF, OM Blaes, 1987, 225, 677). 
However, a small radial motion was seen to damp out the perturbation [148].
% (Blaes, OM; MNRAS, 216, 1985, 799). 
More realistic numerical simulation of advective thick disks 
show that these disks are stable under axisymmetric and non-axisymmetric
instabilities [149].

A frustrating experience of all the model builders had been to 
pinpoint the exact cause of viscosity in an accretion disk which is
supposed to transport the angular momentum radially so that matter may  
fall in. Since radiative and molecular (and ionic) viscosities 
may be very small, especially for radiation pressure supported disks,
one was forced to look for anomalous sources, such as magnetic turbulence [150],
%(Coroniti, 1981), 
non-axisymmetric instabilities [146], %(Papaloizou and Pringle, 1984), 
spiral shock waves [151] % (Sawada, K., Matsuda, T.  and Hachisu, I. 1986; Kaisig, M., AandA, 218, 102, 1989;
% See also  Spruit, 1988 [H, 1988, AandA, 194, 319] for theoretical
% considerations.) 
and amplification of magnetic fields
due to shear instabilities [152]. % (Balbus and Hawley, 1991).
All these models essentially try to obtain a more realistic value of
size of the interacting units ({\it e.g.}, turbulent cells vis-a-vis ions) and
mean velocity of these units ({\it e.g.}, turbulent velocity vis-a-vis mean
thermal velocity). For instance, in Coroniti's model, magnetic flux tubes
are sheared and reconnected repeatedly by the differentially rotating disks
and each of the components separated by reconnection acts as a transporting 
agent. At spiral shocks, the normal component
of the flow is squeezed while the tangential component remains unchanged
thereby bending the flow direction. This process repeatedly 
removes angular momentum. In the case of the study of magnetic amplification
made by Balbus and Hawley, an initially vertical field when perturbed
radially would tend to bend rapidly due to differential motion and 
amplify due to shear. This process has been shown to grow in dynamical timescale.
Recent simulations indicate that the resulting transport 
may not be in outward direction alone, and as a result the
viscous mechanism may not be very efficient.  All these processes indicate 
that $\alpha \lsim 0.01$. In the advective 
disk paradigm (discussed later) solution topologies depend 
very strongly on the viscosity parameter, particularly for such low
viscosity shocks may form if other conditions are satisfied. 
 
There were tremendous progress in the study of jet 
formation in this decade. These studies generally concentrated 
on the wind type outflow from the funnel of thick accretion disks [153]
%(Wiita, Kapahi and Saikia, 1982, 10, 310; Fukue, 1982, Chakrabarti 1986) 
or by magnetohydrodynamic process from all over the disk [154],
%(Blandford and Payne 1982), 
or due to hydromagnetic energy extraction from black holes [155].
%(Blandford and Znajek, 1978), 
(For a contemporary review in jet formation from thick disks 
see, [156].)
% Begelman, Blandford and Rees, 1984, Rev. Mod.Phys.***.). 
Numerical simulations of jets 
(not from disks) were also carried out with high speed super-computers 
to understand how the jets are collimated, how they interact 
with the surroundings, and under what condition cocoons, internal shocks, 
etc. are produced [157]. 
% (Norman, M.L., Winkler, K.-H. A and Smith, M.D., 1982 AandA 113, 285). 
At some point it was believed that these internal shocks
which manifest themselves as hot spots (in M87, for instance),
could be due to periodic outbursts as in dwarf novae [158].
%(DNC Lin, GA Shields, ApJ, 1986, 305, 28). 
The explanations of the interesting behaviour 
at the interface of the cocoon and the jet usually lie in the 
combinations of the Kelvin-Helmholz and  Rayleigh-Taylor instabilities [157-159].
%(Chakrabarti, 1988,  MNRAS 235, 35). 
Jet production {\it from accretion disks} which 
includes radiative transfer was also studied by Eggum, Coroniti and Katz  [160].
%(G;F; J; 1985,  ApJ, 298, 41). 
Among many important results, they found that $80$ percent of the 
energy is trapped and advected to the black hole and 
around $0.4$ percent of the inflowing matter 
actually participates in jets.  Shields, Mckee, Lin and Begelman [161]
%(GA; CF; DNC; MC;ApJ, 1986, 306, 90)
considered wind formation throughout the disk and as long
as the the outflow rate is increased with the inflow rate 
and in fact larger compared to the inflow rate, they found that
the X-ray luminosity heats up the corona and can cause instability in the
incoming accretion rate. The source was assumed to be at the centre. 
K\"onigl [162] % (A, 1989, ApJ, 342, 208)
extended  self-similar models of 
Blandford and Payne to include self-consistently the finite thickness 
of the disk itself and successfully produced self-collimated jets. 
Following this, Chakrabarti and Bhaskaran later showed that it is easier to 
produce magnetically collimated jets if the disk itself is sub-Keplerian [163].
Today, study of the properties of 
the jets in microquasars such as GRS1915+105 verify this
that hard states which is dominated by sub-Keplerian
flow close to a black hole also has dominating wind.

Though Lightman and Eardley considered Comptonization to be the
mechanism of production of hard X-rays in the 70s,
more detailed computation of Comptonization applicable to more
general cases came later. Sunyaev and Titarchuk [164] % (R; LG; AA 1980, 86, 121)
argued that this must be emitted due to Comptonization 
from a hot electron cloud. Nature of emitted power-law
distribution as a function of the electron temperature and optical 
depth of the electron cloud was derived. Its angular distribution 
and polarization properties were further extended for various 
emitting geometries [165]. %(ST, 1985, AA, 143, 374). 
Phillips and M\'esz\'aros [166] %(KC; P.; 1986, ApJ, 310, 284) 
showed that disk polarization could 
be as high as $20$ percent with harder radiation having larger 
degree of polarization. Success in explaining 
hard radiation by hot electron clouds is
followed by efforts to identify the `Compton Cloud', the source
of the hot electrons. Burm [167] % (H.; 1986, AA, 165, 120)
considered formation of magnetically active corona above a disk.
Others considered continuous injection of 
hot electrons and and soft photons and showed that hard X-ray 
spectrum in AGNs could be explained by the Comptonization process. They also
included effects of pair production [168].
%(A.A. Zdziarski and A. P. Lightman 1985 {\it Astrophys. J.}  {\bf 294} L79, 
%M . Kusunose and F. Takahara, PASJ, 41, 263, 1989) 
Kazanas and Ellison [169]
considered formation of accretion shock 
supported by pair plasma and the effect on the AGN spectrum was studied by 
Blondin and K\"onigl [170]. If pair remains coupled to accreted 
matter (which is likely in presence of small magnetic field), then 
pair opacity would reduce the effective Eddington luminosity. As a result the
emitted radiation could contribute strongly to the support the shock. 
Blondin and K\"onigl [170] demonstrated this by using a simple 
minded model of  a shock that is mediated by Fermi-accelerated 
relativistic protons and radiation and in which $e^+-e^-$ pair are produced through 
photon-photon collisions. With the launching of EXOSAT in 1983 and GINGA
in 1987, a large amount of data in X-Rays were available, especially from 
Low Mass X-ray binaries (LMXRB). A number of important discoveries were
made, important among them are the Quasi-periodic Oscillations (QPOs) in
in LMXRBs (in both the neutron star and black hole
candidates) and X-ray pulsars. Frantic efforts followed to model these QPOs
of neutron star candidates. Many galactic and extragalactic sources were also
studied [171]. %(e.g., NR White, A Peacock, G. Hasinger, KO Mason, G Manzo,
%BG Taylor, G Branduardi-Raymont, 1986, MNRAS, 218, 129; Makishima, K.,
%The Physics of Accretion onto Compact Objects, Proceedings of a Workshop
%Held in Tenerife, Spain, April, 21-25, 1986. Lecture Notes in Physics, Vol. 266,
%edited K. P. Mason, M. G. Watson, and N. E. White. Springer-Verlag, Berlin
%Heidelberg New York, 1986., p.249; Stella, L; NE White and AN Parmer, ApJ, 324, 363, 1988). 
The observations clearly show that one requires a source of hot electrons
for Compton scattering as indicated by Sunyaev and Titarchuk [164]. A number of 
sources such as Cyg X-1, LMC X-3, A0620-00, which were already identified to be  black hole candidates by  McClintock and Remillard [172]
%(JE; RA; 1986, Apj, 308, 110) 
by dynamical considerations also showed similar variation of spectral properties where    
a blackbody plus a power-law components were needed to fit the data (c.f., [173]
who consider only modified blackbody models to fit the spectra). 
% Mitsuda (K Mitsuda, H. Inoue, K. Koyama, K. Makishima, M. Matsuoka, Y. Ogawara,
%N. Shibazaki, K. Suzuki, and Y. Tanaka,
%PASJ, 36, 741, 1984) 
%and Hanawa 
%[1989]

\subsection {1990s till today: maturity -  era of perfection}

Accretion disk theory since 1990 till today saw perfecting
a new paradigm called the {\it Advective Disk Paradigm}.
The goal was to achieve a single paradigm so that
observations could be explained within a single framework.
There were too many `models' each suitable for a specific purpose, 
or to explain a specific observation. We have not reached to that
stage of the grand, grand unified, i.e., 
THE MOST GENERAL solution (necessarily time-dependent, because
variabilities observed in all the time scales) which would include every possible
physical processes, such as, heating, cooling, pair-production etc.
and yet, carried out in totally general relativistic framework in 
three-dimensions. But there are every indications that `we are getting there'. 
The Advective Disk Paradigm 
is perhaps closest to reality and today all the 
observations are explained by using some or the other predictions
of this paradigm. Being proponents of this paradigm, we shall naturally
spend some time on this.  A number of branches in accretion physics
which started in earlier decades but whose studies were extended to this 
decade have already been discussed earlier 
and will not be repeated here.

Major shift in accretion disk modeling was due to the perception that the
radial motion must be included since the radial velocity becomes 
velocity of light on the horizon. Thus the inertial force is as
important as the pressure gradient term
and centrifugal pressure terms etc. Some work on spherical accretion model
(which is advective any way, but not disk-like)
still continued [174] %(e.g., Park,M, 1990, 354, 83)
partly because it is easier to
handle spherical flows. Nobili, Turolla and Zampieri  [175]
%(L.; R.; L.; 1991, ApJ, 383, 250) 
presented very general relativistic spherical flow in which various heating/cooling
and dissipation were included. Solutions of generalized rotating flow
is clearly needed. In an earlier work [128] the
complete classification of global solutions of an
{\it inviscid}, polytropic transonic flow was presented which showed 
that in some region of the parameter space, the flow will have multiple sonic
points [126]. Within this  
region, there is a sub-class of solutions where Rankine-Hugoniot
shock conditions are satisfied and standing shock waves are formed due to the
centrifugal barrier. Four locations, namely, $x_{si},\  (i=1..4)$ were 
identified where these shocks could formally be possible, but it was
pointed out that only $x_{s2}$ and $x_{s3}$ were important for accretion 
on black holes since the flow has to be supersonic on a black hole horizon 
and $x_{s1}$ could also be important for a neutron star accretion 
while $x_{s4}$ was  purely a formal shock location. 
In Chakrabarti [176-177] %~\cite{ref:ttaf90a, ref:mnr90b}
viscosity was also added and the complete set
of global solutions in isothermal VTFs
with and without shocks, were presented. In the language of 
Shakura-Sunyaev [29] %~\cite{ref:ss73} 
viscosity parameter $\alpha$, it was shown that if
viscosity parameter is less than some critical value $\alpha_{cr}$,
the incoming flow may either have a continuous 
solution passing through the outer sonic point, or, it can have standing shock 
waves at $x_{s3}$ if the flow allows such a solution in accretion.
For $\alpha >\alpha_{cr}$, a standing shock wave at $x_{s2}$
persisted, but the flow now had two continuous solutions --- one
passed through the inner sonic point, and the other through the
outer sonic point. The one passing through the inner sonic point is
clearly slowly moving in most of the regions, and therefore could be
optically thick when accretion rate is large enough. 
The one passing through the outer sonic point becomes 
optically thin at a large distance. A standing shock
can connect these two pieces if conditions are favourable.
Later analytical [178] % (Chakrabarti, 1992, GRG meeting, Buones Aires) 
and numerical works [179, 38, 180] % ~\cite{ref:cm93, ref:cm95, ref:nh94}
showed that $x_{s3}$ is stable (indeed, at attempt of fitting
AGN spectra by including $X_{s3}$ was made at this stage 
Chakrabarti and Wiita [181], % 1992, ApJ, 387, 21), 
Most importantly, these solutions show that they could join (though
not quite smoothly, since viscosity parameter was chosen to be constant) 
with a Keplerian disk at some distances, depending on viscosity and 
angular momentum [176]. % ~\cite{ref:ttaf90a}. 
A two-component advective flow model (TCAF) was constructed with 
higher viscosity flows on the equatorial plane and lower viscosity
flow away from the equatorial plane [182-184, 56]
%(Texas Symposium, 1994; Chakrabarti and Titarchuk, 1995;
%Chakrabarti, 1996GUT, Chakrabarti, 1997). 
It is now widely believed
that an inflow onto the black holes indeed have two components ({\it e.g.},
[185]).
%(Heindl, Smith and Swank, 2001; .
Incidentally, numerical work of Chakrabarti and Molteni [179] 
was the first to  establish that the non-linear 
solutions which included shock waves were really stable 
and were accurately described by 
Chakrabarti [128, 176] solutions. Molteni, Ryu and
Chakrabarti [186] also tested these solutions with different codes.
It is clear that one could use these non-linear solutions to 
benchmark one's codes to check if the numerical code is dissipative or not. 

In the mid 90s, some hope was raised by Narayan and his collaborators
that a self-similar solution of the same equations used
since 1980s may have a `new' solution [54].  These were termed 
as Advection Dominated Accretion Flow (ADAF)
which apparently advects away most of the energies of the disk. 
In order to investigate if ADAF really represents any new solution branch,
Chakrabarti [183] %~\cite{ref:gut96a} 
found that even when the 
isothermality condition is dropped, the flow topologies 
remained the same as in [176. 177].
%~\cite{ref:ttaf90a, ref:mnr90b}. 
The behaviour of the solution with viscosity was found to be intriguing. 
At very high viscosities
the flow does not have shocks, and sub-Keplerian, optically thin solutions emerge
out of the Keplerian disks, very similar to the Ichimaru [53] %~\cite{ref:i77} 
solution. However, when the
viscosity is intermediate, the solution passes through a standing shock for some region
of the parameter space when $\gamma < 1.5$. It became clear that ADAF does not
represent any new solution, and is, in fact, only a special case of the transonic/advective
disk solutions. ADAF's criterion was that it should be radiatively inefficient.
In that respect, by definition all the solutions of Chakrabarti [128] are 
faithfully  one hundred percent 
ADAF since energy is conserved till the horizon and the entropy generated at the shock
is also advected into the black hole.
Naturally, some confusions followed. Chakrabarti [187-188] wrote more
detail why ADAFs are not only not new, they are incorrect solutions
(as obtained in [54]) as well. Actually those who could not find 
solutions with shocks, did not look for them.
For instance, workers like Peitz and Appl [189] %(J; S; MNRAS 1997, 286, 681), 
did not even look
for shock solutions. 
On the contrary, those who looked for this uniquely stable and most relevant solution found
them without any problem [{\it e.g.}, 180, 190-191]
% 180 Nobuto and Hanawa, 1994, PASJ, 46, 257; 
%: 190 Yang R; and Kafatos, M; AA 1995, 238,  Lu JF; PASJ, 49, 525, 1997, 
%Lu, JF; Yu, KN; Yuan F., Young, ECM,
%1997, AA, 321, 665; TOTH, REFERE etc., Caditz, DM; Tsuruta, S. ApJ, 1998, 501, 242). 
%191 LU et al 1999 PAPER about carpet bombing; 
On the one hand Narayan writes [192]
%In Narayan (1996, Accretion Phenomenon and related outflows, IAU colloquium,  163,
%D.T. Wickramsinghe, G.V. Bicknell and L. Ferrario, p. 79:
" The Global Solutions described in \S 2. are free of shocks. However, Chakrabarti
and his collaborators (cf. Chakrabarti and Titarchuk, 1995) have claimed,
that shocks are generic to ADAFS. Initially Chakrabarti (1990) considered
viscous flows under isothermal conditions, but more recently, with increased
interest in ADAFs, he has switched his attention to adiabatic flows. The clearest
accounts of his results are found in
Chakrabarti and Titarchuk (1995) where the authors
claim (i) low $\alpha$ flows ($\alpha \lsim 0.01$) have shocks and high $\alpha
\gsim 0.01$) flows do not and (ii) low $\alpha$ flows have sub-Keplerian
rotation at large radii while high $\alpha$ flows are Keplerian
at all radii except very close to the black holes. Neither statement
is confirmed in the work described in \S 2.4. In particular, no shocks
are seen in any global solutions, which span a wide range of 
parameter values: $\alpha$ ranging from $10^{-3}$ to $0.3$, and $\gamma$ ranging from
$4/3$ to $5/3$.  What is the source of the discrepancy? Narayan et al. (1997a) suggested that 
it may lie in different philosophies regarding angular momentum parameter
$j$ in equation (2.3). Chakrabarti simply assigns a value to $j$ (in fact, different 
values in different papers) whereas this parameter ought to be treated
as an eigenvalue and determined self-consistently through boundary conditions."
On the other hand, it was shown [193, 194]
%(Ju-Fu Lu, Gu, Wie-Min, Yuan Feng, 1999, 523, 340, ApJ),
that indeed Chakrabarti's solutions are not only correct, correct ADAF solution is
recoverable from it with lesser difficulty if Chakrabarti's method was followed.
For instance [193] writes
`We numerically solve the set of dynamical equations describing advection-dominated
flows (ADAFs) around black holes, using a method to similar to Chakrabarti.
... . We recover the ADAF-thin disk solutions constructed by Narayan, Kato and Honma
in a paper representative of previous works on global ADAF solutions, ... .
Chakrabarti and his collaborators introduced a very clever procedure ({\it e.g.}, Chakrabarti, 1996a) .
The difficulty of finding eigenvalues was simply avoided: $R_s$
and j were to be free-parameters ... Very recently, 
adopting a procedure very similar to Chakrabarti's, Igumenshchev
et al. (1998) obtained global ADAF solutions .... 
By varying the values of two free parameters $R_s$ and $j$,
we can find all the possible solutions:
although some of the solutions
constructed this way may not be physically acceptable, no physical
solutions will be missed if such a `carpet bombing' approach is used."
and [194] writes
%Igumenshchev, IV; MA Abramowicz, MA; Novikov, ID; (MNRAS, 298, 1069), write :
`To avoid this difficulty, Chakrabarti and his collaborators
(see Chakrabarti, 1996 and references therein) introduced 
a very clever mathematical trick. They assumed a different set of boundary
conditions, which are not only at the outer boundary,
but also at the sonic radius, and near to the black hole horizon. In this
way, the most difficult  part of the problem -- finding the eigenvalue --
is trivially solved'. The lesson is therefore ADAF cannot be anything 
except a special case of advective/transonic solutions [176].  
Paczy\'nski, a veteran from the thick accretion disk era, also wonders in 1998, if ADAFs
were new solutions [195]: ``The purpose of this paper is to present
a toy model of a disk accreting onto a black hole
but not radiating, i.e., advection
dominated, presented in the spirit of the 
early 1980s which seems to be 
simpler and more transparent than the spirit 
dominating the late 1990s." Though non-advecting thick disk models were
un-physical, the solutions with constant energy indeed existed before
ADAF, after all. Recent efforts are on include various effects 
into ADAF (see, {\it e.g.}, [196, 197]) to bring it back to known advective solutions.
%196 Park and Ostriker
% 1997 Park and Ostriker
For instance, advective disks (not ADAFs)  always had funnels,
otherwise  the flow would be unstable. This is pointed out
recently [14, 196, 197]. 

So far, there is no observation which suggests that a new solution
other what is in the advective disk paradigm, was required
to explain it. The only observation [131]
that the luminosity of black hole candidates
should be lower compared to that of the neutron stars,
which ADAFs claim to be in their support is by definition a triviality: 
there were black hole solutions with a $100$ per cent advection of
energy [128] and thus in principle, for a given accretion rate
the ratio of these luminosities could be as close to zero as
possible. There is no limit! Today this new terminology  has created 
a major confusion in the subject, and any time a solution has a radial term
(whatever be its accretion rate!) it is termed as an ADAF, even though 
it was proven that ADAF, with original sense of Narayan and Yi [54]
is unstable unless accretion rate in excruciatingly low (basically no accretion).
Today, workers lost track of whether a solution should be called a transonic/advective
flow or simply ADAF (reference too many to cite) so that both
are blissfully cited. Original proponents still think it is a new solution
and stick to the definition of ADAF as the shock-free solutions of Chakrabarti [176].
Indeed readers will miss nothing in the subject if they ignore ADAF-ZDAF altogether.

Extensive numerical simulations of quasi-spherical, inviscid,
adiabatic accretion flows [179, 186, 198-201] %~\cite{ref:mlc94, ref:mrc96, ref:rcm97}
show that shocks form very close to the location
where vertically averaged model of adiabatic flows predict them [128]. The flow 
energy is conserved and the entropy generated at the shock is totally advected 
into the black hole allowing the flow to pass through the inner sonic point. 
Flows with positive energy and higher entropy form supersonic winds.
In presence of viscosity also, very little energy radiates away 
if the accretion rate is low.
Having satisfied with the stability of these solutions
[38, 179, 198-201]
%~\cite{ref:cm93, ref:mlc94, ref:cm95,
%ref:mrc96, ref:lmc98}, 
a unified scheme of advective  accretion disks 
was proposed [38, 202] % ~\cite{ref:unam93, ref:cm95} 
which combined the physics of formation of sub-Keplerian disks with and without shock
waves depending on viscosity parameters and angular momentum at the
inner edge. In this review on historical developments
it is worth quoting from [202]: ``This findings are very significant
as they propose a unifying view of the accretion disks. This incorporates two extreme
disk models into a single framework: for inviscid disks, strong shocks are
produced, and for disks with high enough viscosity, the stable shock
disappears altogether and angular momentum distribution can become Keplerian." 
The solutions remained equally valid for both the black hole and neutron 
star accretions as long as appropriate inner boundary conditions are employed.
A large number of numerical simulations with various codes (mostly
in disguise of ADAF solution) have since then been done but there 
is no new result. However, very recently, a large number of numerical simulation results 
have been reported by Hawley and collaborators [203]
%(e.g., Hawley, 2000, ApJ, 528, 462 and references therein) 
where magnetic fields have been included. In an axisymmetric `cylindrical disk' 
the viscous stress is even found to be close to $0.1-0.2$.

Having advanced a great deal in the subject over last two decades with Newtonian
or Pseudo-Newtonian models, it was time to pause and verify major results 
in Kerr geometry. Prescription to study Keplerian disks was presented 
by Rieffert and Harold [204] %(ApJ, 450, 508, 1995) 
and detailed structures were computed using self-consistent 
vertical height by  [205]. %T. D\"orrer, H. Riffert, R. Stauber, H. Ruder; AA, 311, 69, 1996).
Study of transonic flows and standing shocks 
in Kerr geometry was made in Chakrabarti [206]. % (1996, ApJ, MNRAS).
Two dimensional solutions of advective disks in Kerr geometry 
also showed the presence of funnels along the axis [207] %~\cite{ref:apj96b}
very similar to the thick accretion disks.
Some other advective solutions with  relativistic equation of state
in a different range in parameters were obtained by Gammie and Popham [208].
%(CF; R;  ApJ, 1997, 498, 313). 
Such funnels were non-existent in ADAF solutions
(see, Chakrabarti [14, 188] 
for comparison of solutions.). %~\cite{ref:oebhc98} 
Similar conclusions, that the flow cannot maintain its structure
(from the radiative transfer point of view)
along the pole has been reached by others [195-197]. 
%by Paczy\'nski~\cite{ref:p98} and Park and Ostriker~\cite{ref:po99} as well.
They showed that there is no 
way matter can be kept hot along the pole and matter must cool down. 
Indeed, in  [208] 
%Chakrabarti, (, ApJ, SOK-NOSOK paper) 
it was shown that a single component so-called ADAF flow becomes cooled by the soft-photons
from a Keplerian disk of the same accretion rate. That is why for most of the
explanations of the observations of black hole candidates one requires two components [56].

Several other developments were taking place in the theory of advective flows
particularly in the MHD limit and when the flow is non-axisymmetric.
Chakrabarti [209] % ~\cite{ref:wd90c} 
found all the global solutions of Weber-Devis type of
equations. In this case, the nature of the magnetic field is predefined to be
radial and azimuthal, but nevertheless the solutions indicated new types of
sonic points in both accretion and winds. Instead of three sonic points
(two X-types and one `O' type), one obtains five sonic points (three X-type: slow
magnetosonic, Alfv\'enic and fast magnetosonic and two `O' type). Shocks were 
also found analytically for the first time. Since then similar solutions
have been found for cold flows where pressure and gravity effects were ignored
(see, [210] and subsequent works of this group).
However, till today, no new solution has been found other than those in [209].

With the advent of detailed and accurate observations which demanded more accurate solutions 
for proper explanations, it became clear that the flow with standing [128] or oscillating
shocks [108]  are more relevant than ever before to describe spectral properties of accreting
black holes. In fact, the property of the Centrifugal pressure
supported Boundary Layer (or, CENBOL for short), be it emission of radiation
or driving winds and outflows is closely linked to what is observed from 
the vicinity of a black hole candidate. We refer to the readers a few reviews 
written to summarize the properties of advective disk paradigm 
[70, 211-213]
%(Chakrabarti, 1996, Phys. Rep; IJP 1998; Pescara, 2000) 
where `basic building blocks'
of all possible solutions (as well as non-solutions which are clearly marked) 
are presented and we do not go into details in them.
In realistic cases, neither the viscosity parameter is constant, nor the accretion rate
remains constant. More importantly, one needs to consider processes such as Coulomb 
coupling between ions and electrons, bremsstrahlung, Comptonizaion etc. Also important 
is the availability of the driving forces (such as thermal pressure, centrifugal, magnetic 
etc.) to form outflows. Depending on these factors, a realistic flow would be made up 
of combinations of these basic building blocks of the inflow and outflow.

One relevant matter in this regard is the ratio of the outflow rate to the 
inflow rate  $R_{\dot m}$. Already in the 80s numerical simulations
were carried out to show that very little matter comes out of the disk.
Kusunose [214] % (M; ApJ, 370, 505, 1991) 
considered two temperature Keplerian
disks and found that in some cases more than $50\%$ mass loss is possible.
In more recent works, Chakrabarti [212, 215] gave analytical estimate of mass
loss rate as a function of the shock compression ratio and
%(IJP98, AA99) 
Das and Chakrabarti [216] carried out this analysis using sonic
point method. They showed that $R_{\dot m}$ 
depend on the strength of the shock, when it forms [212, 215]. 
%~\cite{ref:ijpc98, ref:aac99}.
One could argue that in soft states, when the accretion rate is 
high, the inner sub-Keplerian region is cooled down [56]
%~\cite{ref:ct95} 
and the shock is as good
as non-existent and there is no driving force to produce outflows [217]. %~\cite{ref:yati99}.
In the hard states, on the contrary, when the accretion rate is low, the stronger shocks
also cause a very low outflow rate. In the absence of shocks, driving force 
is weak, and the outflow may be negligible (in the absence of strong magnetic fields).
For intermediate shock strength, the outflow rate is high and it could be cooled
down by soft photons from the pre-shock Keplerian disk. Once the flow is cooled down,
it would be super-sonic and fly away but the terminal velocity would be low.
This process may take place repeatedly [217, 218] %~\cite{ref:yati99, ref:cm00}
and cause low frequency variations of the light curve. In this so called `flare' state,
the outflow would be `blobby'. These works therefore directly indicated that 
there should be direct relation of the outflow rate with the spectral states.
As far as the radiative acceleration of the outflows
goes, the situation is far from clear. On the one hand
radiation from the CENBOL is bound to deposit momentum on the outflow, but quick acceleration
will also cause it to slow down due to radiation drag effect. Acceleration 
of particles are difficult from a Keplerian Disk  [219] 
%(Icke, 1980,
%Astro. J. 85(3), 329; Fukue, Kato, Matsumoto, 1985, PASJ, 37, 383;
%Fukue, PASJ, 48, 631), 
but could be a bit more efficient when the photon 
is focussed, for instance, from the CENBOL [220] of an advective disk. 
% (Chattopadhyay and Chakrabarti, 2001, %Int J. Mod. Phys. ****).

Two major types of oscillations of the centrifugal pressure supported boundary layer is 
discussed in the literature, and it is possible that both are important. In one
case the steady shock solution is absent even when the inner sonic point exists
[200]. %~\cite{ref:rcm97}.
In this case, the flow searches for a steady solution by 
first generating entropy through turbulence and then trying to 
pass through the inner sonic point 
non-existent in a steady flow. In another type [108] 
the steady shock solution exists but the cooling time 
in the post-shock/corona region roughly agrees with the infall time in this region.
An oscillation of similar kind is seen at the transition radius between Keplerian
and sub-Keplerian flows as some matter is removed as winds from this 
region [187, 188, 221]. %~\cite{ref:cetal96, ref:korea96}. 
This is present even when no shocks are produced.
A fourth type of oscillation may be set in due to the quasi-periodic cooling of the   
outflow (provided it is high as is the case for average shock strengths $R\sim 2.5-3$
see, [212, 215]) due to enhanced interception of the soft-photons emitted from a Keplerian disk. 
It can be assumed that the quasi-periodic oscillation (QPO) in X-rays observed 
in black hole candidates is the result of one or more of 
these different types of oscillations since other types of oscillations, such as those due to
trapped oscillations [106] %~\cite{ref:khm88} 
or disko-seismology [103] % ~\cite{ref:nvk93}
are incapable of modulating X-rays with large scale amplitude. 
Recently, Molteni et al. [222] %(ApJL, Dec 10th, 2001) 
have discovered, through yet another 
numerical simulation where matter is injected both from the upper and lower halves, 
that the inflow interacts with the outflow and bends like an `warped' disk. This
has some resemblance with the suggestion of Pringle 
[223] %(1997, MNRAS, 292, 136)
where warping at the outer edge could form due to the interaction with radiation
emitted at the central region.

Up until 7-8 years ago, it was 
believed that the hard radiation is the result of 
Comptonization of soft photons by `Compton Clouds' floating around 
the disk or by hot corona. Haardt and Maraschi [224] %(1991, 1993) 
considered 
two phase accretion in AGNs, one with cold component along the equatorial plane, 
and the other with hot component above the cooler component. 
Chakrabarti and Titarchuk [56] for the first time
pointed out that the so-called `Compton Cloud' is the 
{\it sub-Keplerian} inner edge of the disk itself which 
is puffed up due to heat in CENBOL! Here
the higher viscosity Keplerian disk flows on the equatorial plane
inside a sub-Keplerian flow which may have standing or oscillating shock waves
(This `two component advective flow' [TCAF] model
is to be contrasted with the earlier models [47, 53] %~\cite{ref:i77, ref:sle76}
which considered only the Keplerian disks.)
Thus, a new paradigm of accretion disks, based on actual solution
of advective flows, started. Today, this picture is universally adopted in most of the 
models of accretion flows. In the hard state, Keplerian flow rate 
could be very small while the sub-Keplerian
rate could be very high [56]. In the soft state, it 
is the opposite. The de-segregation of matter into 
these two types of rates are believed to be due to the fact that
the flow closer to the equatorial plane is likely to be 
more viscous and with larger Shakura-Sunyaev parameter 
$\alpha$, and therefore is likely to be Keplerian. However,
flows away from the equatorial plane may have smaller 
$\alpha$ and therefore they deviate from a Keplerian disk 
farther out [70, 183, 211]. Contribution to this sub-Keplerian
flow may also have come from wind accretion. This sub-Keplerian 
flow may form shocks at around $10-20R_g$. In the soft state, 
shocks may be nominally present, but would be cooler due 
to Comptonization, and would be as good as non-existent. When the viscosity 
and flow accretion rate is large, the sub-Keplerian region shrinks
as it is cooled down by thermal Comptonization. Only the bulk motion 
Comptonization will take place in flows 
at around $1-3R_g$ thereby producing hard tails in soft states [56, 225],
%Titarchuk, Mastichiadis, A; and Kylafis, N; 1997, ApJ, 487, 834 }.
Outflow also softens the spectra mimicking those from bulk motion
Comptonization [226], %~\cite{ref:c98rap}, 
but as discussed above, it is unlikely 
that there would be strong outflows in the soft state due to the lack of driving forces. 
Now-a-days there are very strong evidences in black hole
candidates such as GRS1915+105 that during outflows
spectral softening would take place and when some of the failed outflow
falls back, the spectrum is hardened [227]. %(Chakrabarti, et al, 2001 MG9).

Similar to Quasars, observers early in this decade
discovered that some of the galactic black hole candidates such as
GRS 1915+105 also have relativistically moving outflows [228] which 
were termed as micro-quasars.
%Mirabel Rodriguez 1994
A shift in understanding that the outflow may actually be coming from 
a small boundary layer (CENBOL) of a black hole rather than from all
over the disk got observational supports from variability data of this
microquasar where it is seen that radio variabilities in the jet are
strongly correlated with that of the Comptonising region. In active galaxies,
even the direct observations [229]
%Junor, Biretta and Livio (Nature, REFEREF) 
showed that the base of the jet in M87 is very narrow,
only a few tens of the Schwarzschild radii. In certain epochs,
when the anti-correlation was observed in the X-ray and 
radio intensities in the black hole candidate GRS 1915+105
such concept of the localization of the jet formation was first proposed
[70, 182]. It was suggested that a strong field in a hot gas feels magnetic 
tension (`rubber-band effect') and is contracted catastrophically evacuating 
the post-shock flow, i.e., the inner part of the accretion disk.
Quoting [70]:`` Since the inner part of the accretion disk
could literally disappear by this magnetic process, radio flares should 
accompany reduction of X-ray flux in this objects. Since the physical
process is generic, such processes could also be responsible for the 
formation of jets in active galaxies and similar anti-correlation 
may be  expected, though time delay effects are to be incorporated
for a detailed modeling.'' Recently this has been quantified [230] %in Nandi et al. (2001)
and it was shown that the magnetic flux tubes indeed collapse with Alfven 
speed inside the CENBOL region.  

As discussed, effort is on to understand various phenomena using a single
advective disk paradigm perhaps to `wrap up'. We present here some 
of the observations and show that these can be addressed by the 
advective disk paradigm quite satisfactorily. 

\noindent {\it Sub-Keplerian motion on a large scale}: Since shocks are 
transitions from supersonic to sub-sonic motion, and since supersonic
flows are sub-Keplerian [70, 183], %(~\cite{ref:gut96a, ref:cpr96}), 
any presence of shocks in a disk would indicate sub-Keplerian motions.
Sub-Keplerian flows rotate slower, and velocity predicted from Doppler 
shifted disk emission lines would correspond to a higher 
central mass. Chakrabarti [145] pointed out that the disk around M87
contained spiral shocks and therefore the flow must be sub-Keplerian. As a result, the 
mass of the central object was found to be around $4\times 10^{9} M_\odot$ rather 
than $2 \times 10^9M_\odot$
as predicted by Harms et al. [144] (see, [231] for references) 
with  calculations purely based on Keplerian motion.
The fact that shock-ionization causes the emission processes on M87 disk has been
stressed recently by several others [232]. %(MA Dopita, et al, ApJ, 490, 202, 1997).

\noindent {\it Sub-Keplerian motion on a small scale}: 
It is the usual practice to assume that the inner 
edge of a Keplerian disk extends till $3R_g$,
the marginally stable orbit. However, the advective disk 
models [70, 183] show that the inner edge could extend to
$\sim 10-20 R_g$ where the CENBOL should form. There are 
overwhelming evidence today that this is indeed the case [233]. 
%~\cite{ref:gil97, ref:n99, ref:dmat99}.

\noindent {\it Power-law hard radiation in very high states}: Chakrabarti and Titarchuk [56] 
pointed out that when the accretion rate is relatively high,
the electrons in the sub-Keplerian region become cooler and this region becomes practically
indistinguishable from that of Keplerian disk. However, very close to the black hole horizon,
matter moves with almost velocity of light and deposits its bulk momentum onto the photons
thereby energizing these photons to very high energy forming a power-law. This power-law
is the hall-mark of all the known black holes [234]. %~\cite{ref:bor99}. 
The success of this model
crucially hangs on the transonic flow solution  which utilizes the fact that the inner
boundary condition is independent of the history of incoming matter. 

\noindent{\it Quasi-Periodic Oscillations from black hole candidates}: 
X-rays from galactic black hole candidates often show persistent oscillations
which are quasi-periodic in nature.
Advective disk solutions do allow oscillations especially in the
X-ray emitting regions. The observation of these oscillations during the transition of states
is the triumph of the advective disk model ({\it e.g.}, [235]). %~\cite{ref:rut99}. 
Chakrabarti and Manickam [218] proves that the oscillation occurs 
in hard X-rays only strongly pointing to the shock-oscillation origin, 
as in this model post-region acts as the Comptonising region [56]. 

\noindent{\it Outflows and their effects on the spectral properties}: 
Recent high resolution observations of jets in M87 strongly suggest that
they are produced within a few tens of Schwarzschild radii of the horizon [229], 
%~\cite{ref:jbl99},
strongly rejecting ADAF (and its variants such as ADIOS [236]) models for the outflows,
which has no special length scale at these distances. 
Given an inflow rate, one is now capable of computing the outflow rate
when the compression ratio at the shock surface is provided [212, 215]. %~\cite{ref:bang98, ref:aac99}.
This solution naturally predicts that the outflow must form at the CENBOL. The
Globally complete Inflow-Outflow Solutions (GIOS) were also found [216, 236]. %~\cite{ref:bang98}.
In presence of winds, the spectra is modified: hardening of the soft-state and softening
the hard state is predicted and is observed in GRS1915+105 [227].

\noindent{\it Quiescence states of black holes}: Chakrabarti [188] %~\cite{ref:oebhc98} 
and Das and Chakrabarti [216], %~\cite{ref:daschak}, 
pointed out that in some regions of the
parameter space, the outflow could be so high that it evacuates the disk and forms
what is known as the quiescence states of black holes. A well known example is the
starving black hole at Sgr A* at our galactic centre whose mass is
$\sim 2.6\times 10^6 M_\odot$ and the accretion rate is supposed to
be around $\sim 10^{-5} M_\odot$ yr$^{-1}$ which is much smaller compared to the
Eddington rate. Quiescence states are also seen in stellar mass black hole
candidates such as A0620-00 and V404 Cygni. Another way of producing these states
is to  use well known viscous instability in an accretion disk as used in models of dwarf-novae
outbursts [56, 238]. %(Chakrabarti and Titarchuk, 1995; Cannizzo, JK, ApJ, 494, 366, 1998).

\noindent{\it On and Off-states during QPOs in black holes}: 
The black hole candidates GRS1915+105 displays a variety of behaviour: usual high 
frequency QPOs are frequently interrupted by low frequency oscillations. 
While the QPO frequency can be explained by the shock oscillations, the switching of 
on and off states is explained by the duration in which extended corona becomes 
optically thick. It is possible that outflows are slowed down by this
process and the matter falls back to CENBOL, extending the duration of the `on' state
(which, if exists, is found to be comparable to the duration of the `off' state
[239]). %~\cite{ref:y99, ref:b97}).
The QPO frequency does evolve during this time scale and it is suggested that this is due to
the steady movement of the inner edge of the Keplerian accretion disk [239] %~\cite{ref:tru99}
or the steady movement of the shock itself in viscous time scale [218]. %~\cite{ref:cm99}.
The correlation between the duration of the off-state and the QPO frequency has been
found to agree with the observations [218].

\noindent{\it Quick state transition in GRS 1915+105}: 
Very accurate observations of the galactic black hole candidate GRS 1915+105 brought 
a large number of 
challenges along with it. QPO frequencies changed  everyday in unpredictable
ways. Classification of light curves  were carried out by Belloni et al. [241]
%(2000, AA, 355, 271) 
and subsequently rearranged in the order by Nandi et al.  [242]. % (2000). 
In particular Belloni et al. [241] pointed out that
GRS1915+105 likes to go from a harder state (C) to softer state (B) through another state
(A).  Using the advective disk paradigm, such behaviour could be fully 
understood [243, 227]. So far, no alternative explanations exist
of this curious behaviour.
%(Chakrabarti et al. MG9).

\noindent {\it Relationship of outflows and the black hole states}: 
It has been already pointed out that in advective disks
hard states (shock strength is high) should have smaller
outflow rates than the intermediate states (intermediate shock strength), but there
should be virtually no outflow in the very soft states (shock non-existent).
This picture has gained some support as well 
from observations [244]. %~\cite{ref:f99}.

\noindent {\it Polarization of AGNs}: Several studies of polarization suggests that
in AGN, polarization matches with thick disk models
more accurately than those obtained from the thin disk models [245].
%.  (Coleman, HH; Shields, GA, ApJ, 1990, 363, 415). 
Kartje and K\"onigl [246] % (JF, A, ApJ, 1991, 375, 69) 
found that due to multiple reflection on the funnel wall polarization 
could be as high as $10\%$ and suggested that polarization of
X-ray selected BL Lac objects  may be due to this disk geometry.
A post-shock flow is almost as good as a thick disk, only 
better since the radial motion is included. Even without shocks,
funnel walls are produced in advective disks of weaker viscosity. Thus
perhaps signatures of such disks have been observed by this method.

The 90s also saw a few major modifications of the standard
accretion disk model itself. Even though inner disk needed 
major modifications by introduction of the radial motion,
the Keplerian disk is still a major source soft photons in the 
spectrum from AGNs and X-Ray binaries and such improvements
were thus quite natural and welcome additions. For instance,
Wandel and Liang [247] % (A; EP; ApJ, 380, 84, 1991) 
gave analytical solutions for two distinct states of a Keplerian
disk, one being cool ($\sim 10^6$K) and other hot ($\sim 10^9$K). 
Both single temperature and two temperature solutions 
were considered. Another study was to compute spectral hardening
factor $f$. If electron scattering is the dominating 
opacity source and if the X-Ray spectrum is affected by Comptonization
the local flux in the X-ray range is expected to form a 
Wien peak whose intensity is approximated by a 
diluted black body $F_\nu \sim \pi B_\nu (f T_{eff})/f^4$
where $\nu$, $T_{eff}$, $f$, and $B_\nu$ are the 
frequency, effective temperature, spectral hardening factor
and Planck function respectively.
This work was extended to include transonic disks as well [248].
%(Luo, C; and Liang, EP; ApJ, 498, 307).

Shimura and Takahara  [249] % (T; F; ApJ, 1995, 445, 780) 
showed that the emitted radiation spectrum  from 
a Keplerian disk around a black hole could be
described by a diluted black body spectrum with a
hardening factor $f \sim 1.8-2.0$ for rates close to the
Eddington rate and  $\sim 1.7$ for rates close to 10 percent 
of the Eddington rate. These factors must be taken 
into account while estimating the mass of a black hole.

There was remarkable improvements in understanding
the nature of our Galactic Centre in this decade. Kinematic 
considerations fitting [Ne II] line map showing continuous variation
of velocity along a spiral pattern showed that the 
mass of the black hole at the Center could be around $2 \pm 0.5 \times
10^6 M_\odot$ [250]. % [Lacy, JH, Achtermann, JM, Serabyn, E., ApJ, 380, L71, 1991].  
Preliminary fits of its spectrum by dissipating spherical accretion model 
also argued for its mass about this high  [251]. %(Melia, F., ApJ, 387, L25, 1992].
More recent studies of motion of stars around the
center yields the estimate of the central mass to 
be converging to $2.65 \times 10^6M_\odot$ [252]. % (Ghez et al, OEBU)
From polarization measurements any  of the 
ADAF solutions has been completely ruled out [253].
%(Agol, E. 2000ApJ...538L.121). 
It seems to be a clear case of an advective disk 
with a very small accretion rate. 

It was pointed out [254, 255] %~\cite{ref:grav93, ref:cprd96} 
that since advective disks close to a black hole need not be Keplerian, it would affect the
gravitational wave properties of a coalescing binary. The angular
momentum loss to the disk by the companion in comparison to the angular momentum
loss to gravitation wave was found to be very significant [255]. %. ~\cite{ref:cprd96}.
The variation of the signal is shown in [212]. % ~\cite{ref:ijpc98}
One of the exciting predictions of this scenario is that since the spectrum of an
accretion disk contains a large number of informations ({\it e.g.} mass of the central black holes,
distance of the black hole, accretion rate, and viscosity parameter) a simultaneous
observation of the electromagnetic spectrum from the disk and the gravity wave spectrum
(which also must depend on those parameters, except possibly the distance)
should tighten the parameters very strongly. 
Recently, ADAF model of  extremely low accretion rate was used to repeat
these computations, and not-surprisingly, no significant change
in gravitational signal was found [256]. % ~\cite{ref:n99?}. 
If little matter is accreting, it is as good as having no accretion disk at all. ADAF model ideally
valid for zero accretion rate systems and not found to be true in any of
the  realistic accreting systems. As discussed earlier,
evidences of the oscillating shocks, bulk motions etc. (which are absent in ADAF)
are abundant. Thus it is likely that the gravitational signal
{\it would be} affected exactly in a way computed by [255]. % ~\cite{ref:cprd96}

In the  80s iron lines were observed in galactic and extra-galactic sources [171].
% (NR White, A Peacock, G. Hasinger, KO Mason, G Manzo,
%BG Taylor, G Branduardi-Raymont, 1986, MNRAS, 218, 129; Makishima, K.,
%The Physics of Accretion onto Compact Objects, Proceedings of a Workshop
%Held in Tenerife, Spain, April, 21-25, 1986. Lecture Notes in Physics, Vol. 266,
%edited K. P. Mason, M. G. Watson, and N. E. White. Springer-Verlag, Berlin
%Heidelberg New York, 1986., p.249; Stella, L; NE White and AN Parmer, ApJ, 324, 363, 1988). 
The observers also interpreted that the iron-line may be emitted from the inner 
part of the accretion disk, where the Shakura-Sunyaev disk may be cooler [257].
They 
%Matt, Fabian and Ross (G; AC; RR, 1993, MNRAS, 262, 179)
consider shining cooler disk of high mass accretion rate with 
extended sources of X-rays which photo-ionize cooler disk in order to
produce observed equivalent width of several hundreds of electron volts.
Recently one of the wings has been seen to have been strongly red-shifted
giving impression that the emission is extremely close to a black hole.
Details of computation of the nature of the line profile is in [258].
%(Fanton, C.; Calvani, M. de Felice, F., Cadez, A; PASJ, 49, 159).
While it is unclear if disks around a stellar mass black hole
could be as cooler as is required for these lines, in AGNs the
prospect is better, where variation of these lines 
can give informations about inhomogeneities of the disk. 
Placing the X-ray source to excite the lines is a major problem.
Some suggestions as to how the correlation of short time-scale X-ray variabilities
with variation of iron line emission could be used to infer the mass and spin of 
AGNs are given by  Reynolds et al. [259]. 
The interpretation of disk origin of the iron-line is not unique however, since in the modern
understanding of the accretion disk model, standard Keplerian disk very often does
not extend to the marginally stable orbit, but is truncated farther out.
It is believed that one could produce iron-line in outflows as well [56].

Resonance lines of iron have been seen in several black hole candidates  [260, 261].
%(Ebisawa, et al. 1996, Oosterbroeck, et al. 1996). 
Usually one of the two observed wings is found to be stretched compared to the
other and it is explained  to be due to the combination of the
Doppler shift and the gravitational red-shift. Generally, it is
difficult to explain very large equivalent width of the lines in this
models.  This problem can be circumvented if the lines are
assumed to be coming from outflowing winds. The stretched wing
would then be due down-scattered emission lines [56]. The idea
of line emissions from the winds is finding supports by other workers
as well [261-263].

\section{Concluding remarks}

It is clear that since a `carpet bombing' method was used, solutions of accretion disks  presented 
in [56], [176] and [183] are the most general ones and the method of obtaining them are the least difficult of
all. All the outstanding issues related to the black hole accretion, namely,
the disk-jet connection, spectra of disks, relatingship between the spectral states and jet formation
quasi-periodic oscillations, variations of the Comptonizing region,
etc. are explained with such solutions or the time-dependent 
solutions ({\it e.g.} [108], [222] etc.). Concerted future efforts should be made to 
understand  the observations on the basic of these general solutions rather than
making new `models' which are at best hand-waiving. To keep up with competition,
observers very often tend to `listen and learn' the basic results from theoretical
solutions and make new models with these salient features without ever referring to the
original theory! Particularly latter has slowed down the progress considerably in recent years
since it is causing confusion especially for the new-comers. 

\section*{Acknowledgments}

The author thanks DST Grant No. SP/S2/K-14/98 for a partial support through the project `Analytical 
and numerical studies of astrophysical flows around black holes and neutron stars'.

 {}

\end{document}